\pdfoutput=1

\documentclass[11pt,twoside,a4paper,cmspaper,final,collab]{cms-tdr}
\def\svnVersion{61292:61370}\def\cmsCernNo{2011-066}\def\cmsCernDate{\today}\def\cmsMessage{Submitted to Physical Review Letters}

\begin{document}\cmsNoteHeader{TOP-10-008}

\hyphenation{had-ron-i-za-tion}
\hyphenation{cal-or-i-me-ter}
\hyphenation{de-vices}
\RCS$Revision: 61370 $
\RCS$HeadURL: svn+ssh://alverson@svn.cern.ch/reps/tdr2/papers/TOP-10-008/branches/prl/TOP-10-008.tex $
\RCS$Id: TOP-10-008.tex 61370 2011-06-15 18:43:45Z alverson $
  \newcommand{\mt}{\ensuremath{M_{l\nu b}}\xspace}
  \newcommand{\cosThetaPol}{\ensuremath{\cos{\theta^*}}\xspace}
  \newcommand{\etalj}{\ensuremath{\eta_{{\rm light~jet}}}\xspace}
  \newcommand{\ipsig}{\ensuremath{{\rm IP}/\sigma_{{\rm IP}}}\xspace}

  \newcommand{\xsec}{83.6\xspace}
  \newcommand{\xsecerr}{29.8\xspace}
  \newcommand{\xseclumi}{3.3\xspace}

  \ifthenelse{\boolean{cms@external}}{\newcommand{\xright}{bottom\xspace}}{\newcommand{\xright}{right\xspace}}
  \ifthenelse{\boolean{cms@external}}{\newcommand{\xleft}{top\xspace}}{\newcommand{\xleft}{left\xspace}}
  \ifthenelse{\boolean{cms@external}}{\newcommand{\xRight}{Bottom\xspace}}{\newcommand{\xRight}{Right\xspace}}
  \ifthenelse{\boolean{cms@external}}{\newcommand{\xLeft}{Top\xspace}}{\newcommand{\xLeft}{Left\xspace}}

\cmsNoteHeader{TOP-10-008} 
\title{\texorpdfstring{Measurement of the $t$-channel single top quark production cross section in \Pp\Pp\ collisions at $\sqrt{s} = 7$~TeV}{Measurement of the t-channel single top quark production cross section in pp collisions at sqrt(s) = 7 TeV}}

\address[neu]{Northeastern University}
\address[fnal]{Fermilab}
\address[cern]{CERN}
\author[cern]{The CMS Collaboration}

\date{\today}

\abstract{
Electroweak production of the top quark is measured in $\Pp\Pp$ collisions at
$\sqrt{s} = 7$~TeV, using a dataset collected with the CMS detector at the LHC and corresponding to an integrated luminosity
of 36~pb$^{-1}$.
With an event selection optimized for $t$-channel production,
two complementary analyses are performed.
The first one exploits the special angular properties of the signal,
together with background estimates from data.
The second approach uses a multivariate analysis technique to probe
the compatibility with signal topology
expected from electroweak top quark production.
The combined measurement of the cross section is
$83.6 \pm 29.8\ ({\rm stat.+syst.}) \pm 3.3\ ({\rm lumi.})$~pb,
consistent with the standard model expectation.
}

\hypersetup{%
pdfauthor={CMS Collaboration},%
pdftitle={Measurement of the t-channel single top quark production cross section in pp collisions at sqrt(s) = 7 TeV},%
pdfsubject={CMS},%
pdfkeywords={CMS, physics}}

\maketitle 

\begin{figure}[hbtp]
	  \begin{center}
	    \includegraphics[width=0.3\columnwidth]{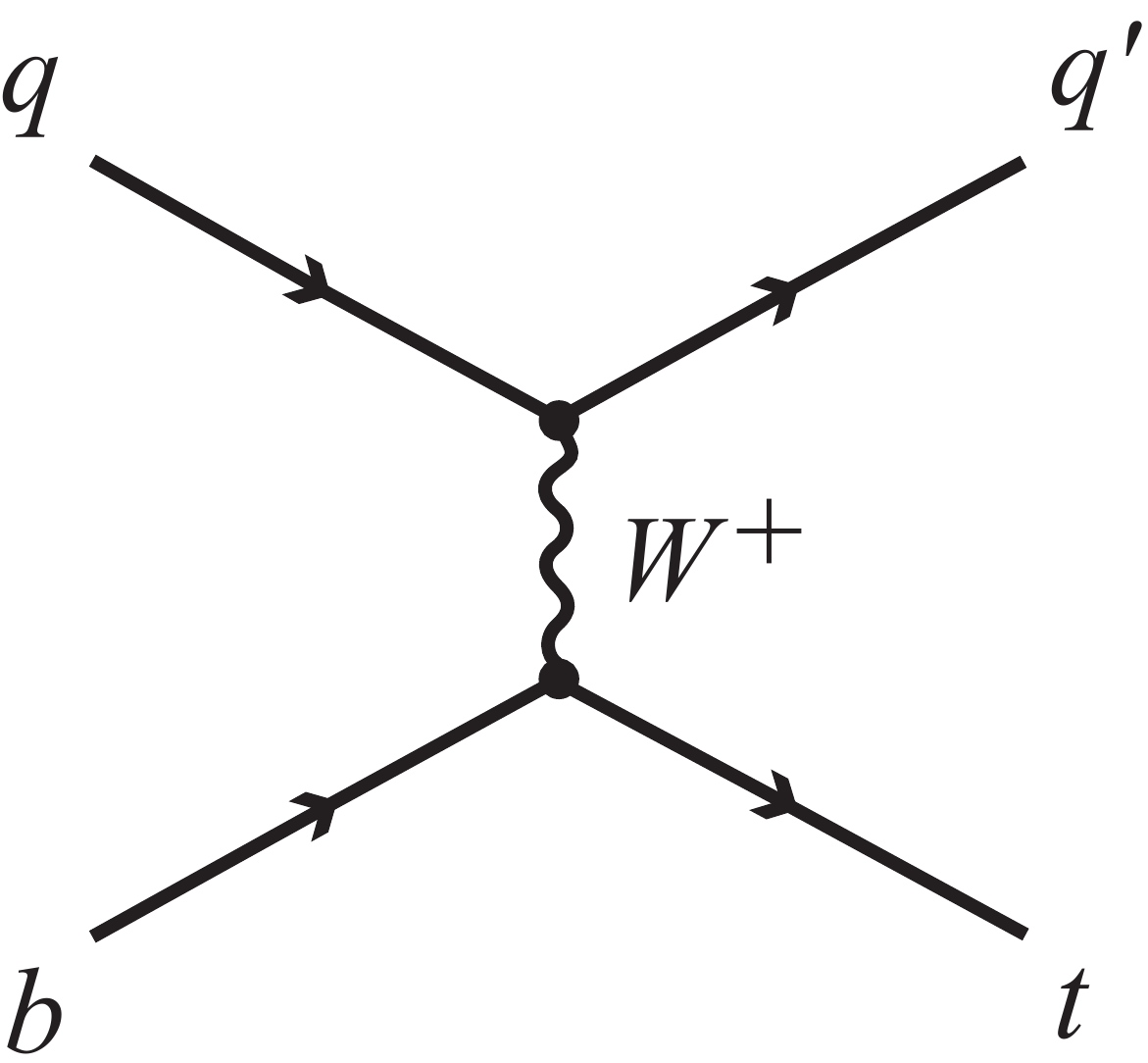}
		\hspace{0.1\columnwidth}
	    \includegraphics[width=0.3\columnwidth]{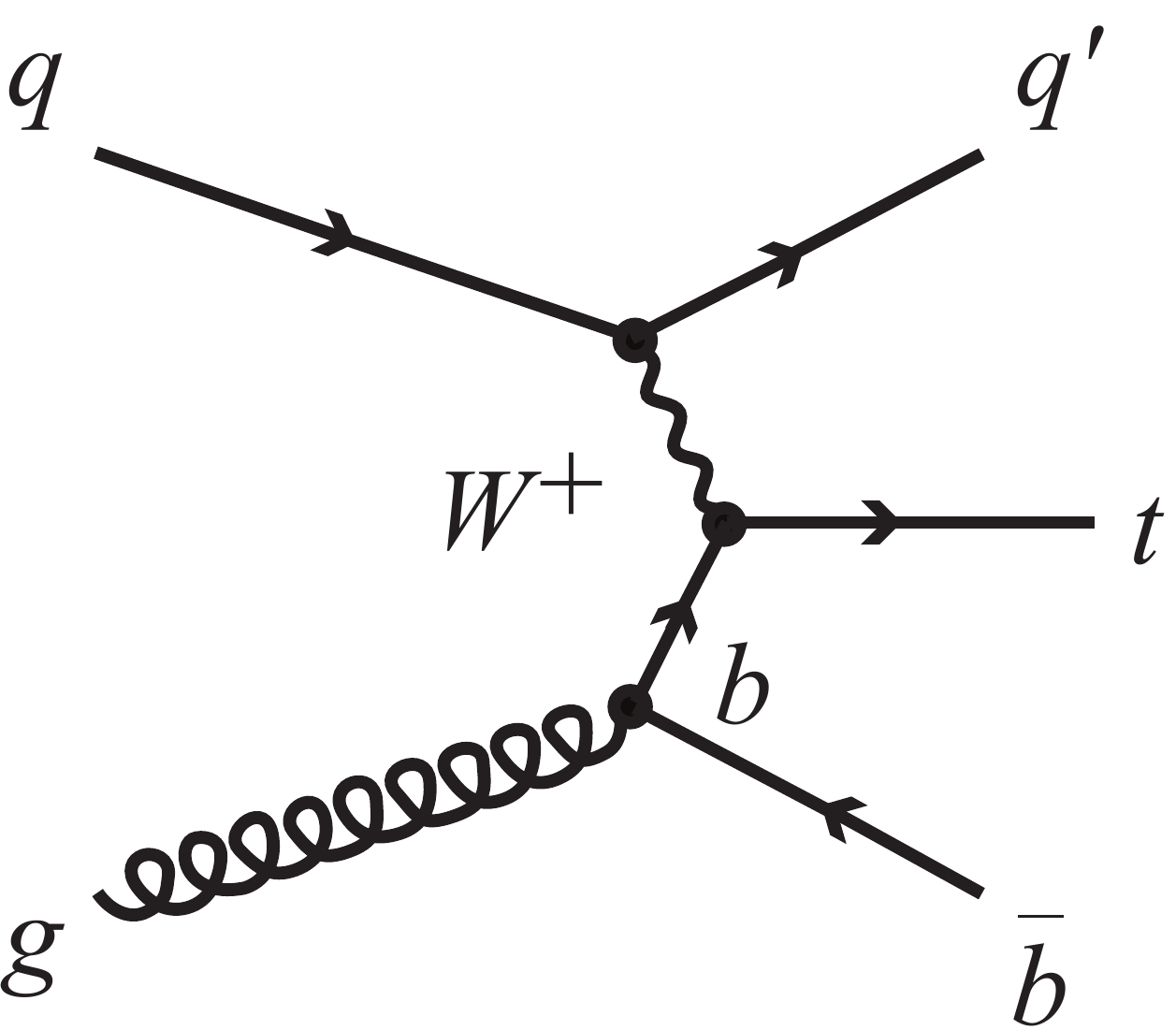}
	    \caption{\label{fig:FG}Feynman diagrams for single top quark production in $t$ channel: $2 \to 2$ (left) and $2 \to 3$ (right) processes.}
	  \end{center}
	\end{figure}

Electroweak theory predicts three mechanisms for single top quark
production in hadron-hadron collisions: $t$-channel, $s$-channel, and $\cPqt\PW$
(or $\PW$-associated) production.
 Single-top events have been observed by the D0
and CDF experiments at the Tevatron $\Pp\Pap$ collider~\cite{Abazov:2009ii,Aaltonen:2009jj,Group:2009qk}, and
first measurements of individual channels have recently been reported~\cite{Abazov:2009pa,Aaltonen:2010jr}.
 In proton-proton collisions at 7~TeV, $t$-channel single top quark
production, Fig.~\ref{fig:FG}, has the largest cross section and
 the cleanest final-state topology, because of the presence of a light jet recoiling against the single top quark.
  Next-to-leading order (NLO) computations with resummation of collinear and soft-gluon corrections at next-to-next-to-leading logarithmic accuracy predict $\sigma_{\cPqt}~=~64.3^{+2.1}_{-0.7}{}^{+1.5}_{-1.7}$~pb~\cite{Kidonakis:2011wy}, for a top mass of $m_{\cPqt}~=~173~\GeVcc$ and with parton distribution functions (PDFs) as given in Ref.~\cite{Martin:2009iq}. The first uncertainty comes from doubling and halving the renormalization and factorization scales and the second from PDF uncertainty at the 90\% confidence level.

This Letter presents the first measurement of the $t$-channel single
top quark production cross section in $\Pp\Pp$ collisions at $\sqrt{s} = 7$~TeV
 in the decay channels ${\cPqt \to \Pe} \cPgn \cPqb$, ${\cPqt \to} \Pgm\cPgn\cPqb$, and ${\cPqt \to} \Pgt \cPgn \cPqb$ with leptonic $\Pgt$ decays.
 Two complementary measurements
are performed. The first analysis exploits
two angular observables sensitive to $t$-channel single top quark
production: the non-central pseudorapidity distribution of the
light jet, and
the cosine of the angle between this jet and the final-state lepton,
in the reconstructed top-quark rest frame.
A multivariate analysis technique with boosted decision trees (BDT)~\cite{decision_trees,boost}
 is used in  the second method, which probes the overall
compatibility of the signal event candidates with the event topology
of electroweak top quark production. Hereafter, these analyses will
be referred to as 2D and BDT analysis, respectively.

Both analyses use a data sample corresponding to an integrated luminosity of 35.9 $\pm$ 1.4~\pbinv~\cite{lumi}, collected by the Compact Muon Solenoid (CMS) detector~\cite{JINST} operating at the Large Hadron Collider (LHC).
 The central feature of the CMS detector is a superconducting solenoid providing a field of 3.8~T. Located within the solenoid are the silicon pixel and strip tracker, the crystal electromagnetic calorimeter and the brass/scintillator hadron calorimeter. Muons are measured in gas-ionisation detectors embedded in the steel return yoke. In addition to the barrel and endcap detectors,
 a quartz-fiber Cherenkov detector extends the jet acceptance to $\vert\eta\vert=5$,
 where the pseudorapidity $\eta$ is defined as $\eta=-\ln{\tan{\frac{\theta}{2}}}$,
where $\theta$ is the polar angle of the particle or jet trajectory
with respect to the counterclockwise beam direction.

Events are selected by requiring the presence of at least one muon or electron having high transverse momentum (\pt).
 The particle flow (PF) algorithm described in~\cite{pflow} performs a global event reconstruction and provides the full list of particles identified as electrons, muons, photons, charged and neutral hadrons.
 A fully reconstructed isolated muon (electron) candidate originating from the leading primary vertex is required~\cite{top-10-002} with \pt $>$ 20 (30)~\GeVc, $|\eta|<2.1$ ($2.5$), and a veto is applied on additional leptons passing lower thresholds.

Jets are reconstructed using the anti-$k_T$ algorithm~\cite{Cacciari:2008gp} with a distance parameter of 0.5, clustering particles identified by the PF algorithm.
 Jets within the full calorimeter acceptance are considered, with \pt $>$ 30~\GeVc
 after corrections for the jet energy scale, as determined from simulations and collision data~\cite{jes}.
 The BDT analysis first identifies isolated leptons,
 which are then excluded form the jet clustering step.
 In the 2D analysis, possible jet-lepton ambiguities are
 resolved on the basis of the distance $\Delta R \equiv \sqrt{(\Delta\eta)^2 + (\Delta\phi)^2}$
 between the reconstructed jet and the nearest lepton.
 The event is accepted for further analysis only if exactly two jets are reconstructed.

 In order to reduce the large background from \PW\ + light partons, we apply a \cPqb-tagging algorithm~\cite{btag} that calculates the signed 3D impact-parameter IP significance (\ipsig) of all the tracks associated with the jet passing tight quality criteria.
 The tracks are ordered decreasingly, following their value of \ipsig,
 and a tight selection threshold is applied on the impact
 parameter significance of the third track in the list. This threshold
 corresponds to a \cPqb-jet identification efficiency of $\sim$40\% and a
 misidentification rate of $\sim$0.1\% determined in data as a function of \pt and $\eta$~\cite{btag}.
 The 2D analysis exploits the expectation that most of the signal events, even in the $2\to 3$ process, have only one \cPqb\ quark inside the tracking acceptance ($|\eta| < 2.4$).
 Events are rejected if the jet failing the tight threshold passes a loose threshold on the IP significance of the second track. The loose threshold corresponds to an efficiency and misidentification rate of about 80\% and 10\%, respectively.
 The BDT analysis applies no veto on the second \cPqb-tagged jet, and rejects events where the jets are back-to-back, which are found to be poorly reproduced by the W + jets simulation.
 To further suppress contributions from processes where the muon (electron) does not come from the decay of a W boson, we require a transverse mass of the W boson $M_T >$ 40 (50)~\GeVcc, where the transverse missing energy (\MET) from the PF algorithm is used as a measurement of the \pt of the undetected neutrino.

The 2D analysis selects 112 (72) events in the muon (electron) decay channel, while the BDT analysis selects 139 (82).
 In both analyses a signal purity of around 18\% (16\%) is expected in the muon (electron) decay channel.
 The main backgrounds are ${\cPqt\cPaqt}$, ${\PW\cPqb\cPaqb}$, \PW\ + light-partons, \PW\cPqc, \cPqt\PW, and processes where the lepton does not originate from a \PW/\cPZ, hereafter called QCD events.

The $t$-channel events from Monte Carlo simulation used in this study have been generated with the \MADGRAPH~4.4 event generator~\cite{madgraph}.
To give a fair approximation of the full next-to-leading order properties of the signal, we combine the dominant NLO contribution (2$\rightarrow$3 diagram ${\cPq\cPg \to \cPq^{\prime}\cPqt\cPaqb}$ and its charge conjugate) with the leading order diagram (2$\rightarrow$2, ${\cPq\cPqb \to \cPq^{\prime}\cPqt}$) by a matching procedure based on Ref.~\cite{boos2000}.
 \MADGRAPH is used also for ${\cPqt\cPaqt}$, single top $s$ and $\cPqt\PW$ channels, and \PW/\cPZ\ + jets.
 The remaining background samples were simulated using \textsc{Pythia}~6.4.22~\cite{pythia}; these include di-boson production ($\PW\PW$, $\PW\cPZ$, $\cPZ\cPZ$), $\cPgg$ + jets, multi-jet QCD enriched in events with electrons or muons coming from the decay of \cPqb\ and \cPqc\ quarks or muons from the decay of long-lived hadrons, and particles with large probability to leave a high-energy electromagnetic deposit.
 The CTEQ~6.6 PDF sets~\cite{cteq6} are used for all simulated samples.
 All generated events undergo a full simulation of the detector response based on \GEANTfour~\cite{geant}.

 The NLO theoretical prediction is used to normalize the single-top production in $s$ and $\cPqt\PW$ channels~\cite{Kidonakis:2010tc,wt} and di-boson processes~\cite{Gavin:2010az}.
 The $\cPqt\cPaqt$ cross section is normalized to 150~pb, with uncertainty constrained to the result of a dedicated analysis. The same analysis constrains the ${VQ\bar Q}$ ($V=\PW,\cPZ$ and $Q=\cPqb,\cPqc$) and $\PW\cPqc$ components, obtaining in particular a factor of 2$\pm$1 for $\PW\cPqb\cPaqb$ over the LO prediction.

The QCD yield is estimated from the same data set by a maximum likelihood fit to the $M_T$ distribution after all other selection criteria have been applied.
 The $M_T$ distribution for QCD events is taken from a control sample obtained by inverting the lepton isolation requirement.
 The latter requirement rejects most of the signal-like events (single top, \PW/\cPZ\ + jets, \ttbar) leaving a QCD-dominated sample. The distribution for the sum of all non-QCD processes is taken from simulation.
 The uncertainty on this estimate is conservatively estimated such as to cover the differences observed when varying the fit range and the QCD shape.

 The BDT analysis normalizes the result of the \PW\ + jets simulation to the inclusive W cross section at NNLO~\cite{Gavin:2010az}, while collision data are used in the 2D analysis to extract the normalization of the \PW\ + light-partons background.
 Two control samples are used, orthogonal to the standard selection. Control sample `region-$A$', dominated by the \PW\ + light partons background, is defined by the requirement of one isolated lepton and exactly two jets, one of which is required to be within the tracker acceptance and with at least two tracks satisfying the quality selection of the \cPqb-tagging algorithm. Both jets should fail the tight \cPqb-tagging selection.
 A second control sample, `region-$B$',
is defined as a subset of the former where at least one jet passes the loose \cPqb-tagging selection although it fails the tight one.
 In both samples a fit of the $M_T$ distribution is performed, allowing the QCD and \PW\ + light-partons background to float, while all other processes,
 including heavy-flavour contributions and the $t$-channel signal, are constrained to their expected values.
 A scale factor of 1.27 in the muon and 1.05 in the electron decay channel is observed between the number of \PW\ + light-partons events obtained from the fit in sample region-$B$ and the predictions from simulation.
 These scale factors are used to obtain the central value of the predicted background.
 A $\pm 30\%$ ($\pm 20\%$) uncertainty is assigned on the muon (electron) scale factor, covering both the statistical uncertainty from the fit, the difference between
the background predictions obtained from the two control samples, region-$A$ and region-$B$,
 and between data and simulation results for both samples.
 The normalization of \cPZ\ + jets background is rescaled by the same factor as that for the \PW\ + light-parton background.

A top-quark candidate is reconstructed in each event by pairing the \cPqb-tagged jet with a \PW-boson candidate. The latter is reconstructed by imposing the \PW-boson mass as a kinematic constraint, leading to a quadratic equation in the longitudinal neutrino momentum, $p_{z,\cPgn}$. When two real solutions are found the smallest $|p_{z,\cPgn}|$ is taken, and for complex solutions the imaginary component is eliminated by modifying $E_{{\rm T,x}}^{\rm{miss}}$ and $E_{{\rm T,y}}^{\rm{miss}}$ independently, such as to give $M_T = M_W$~\cite{julia}.

\begin{figure}[htbp]
\centering
\includegraphics[width=0.7\columnwidth]{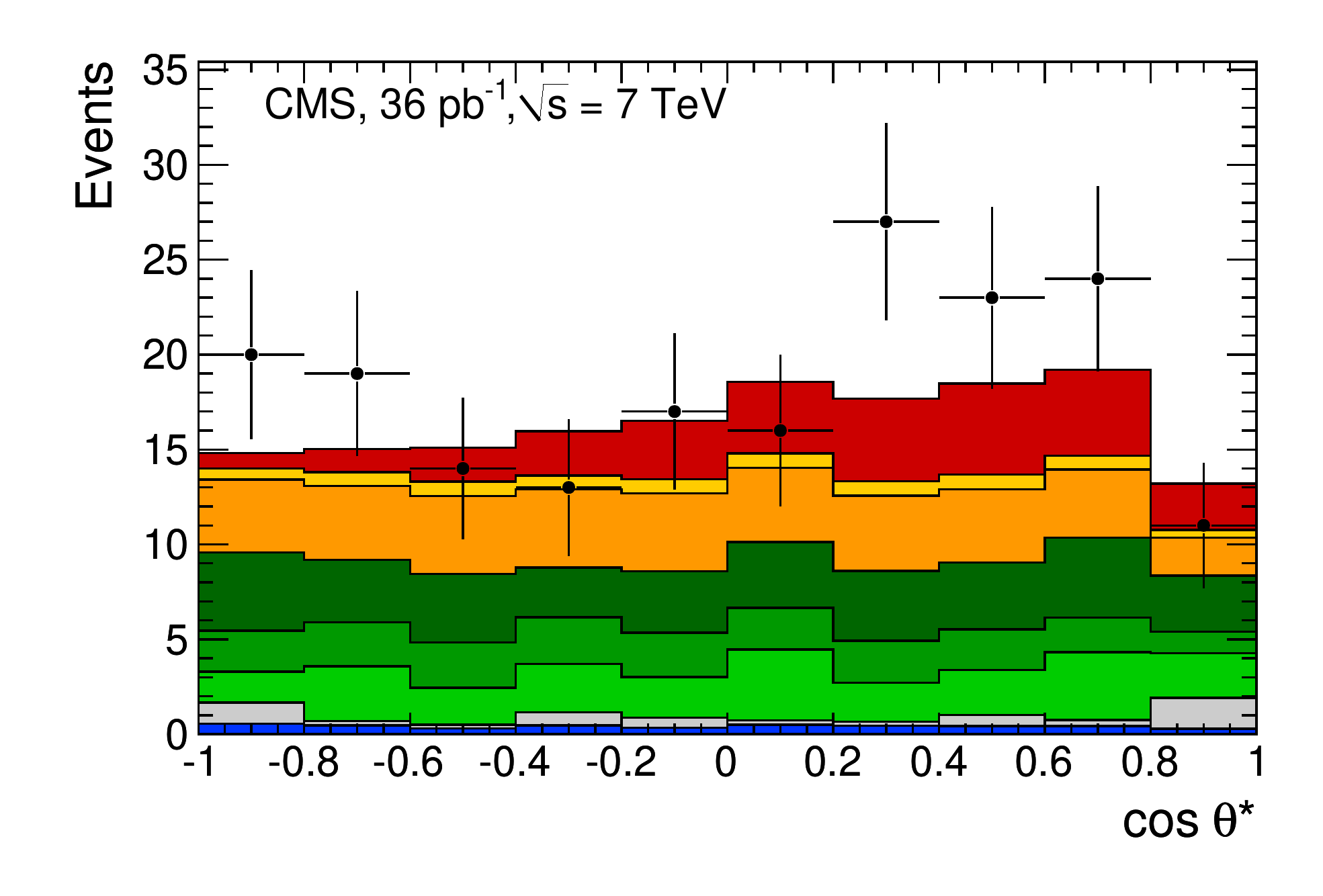}
\includegraphics[width=0.7\columnwidth]{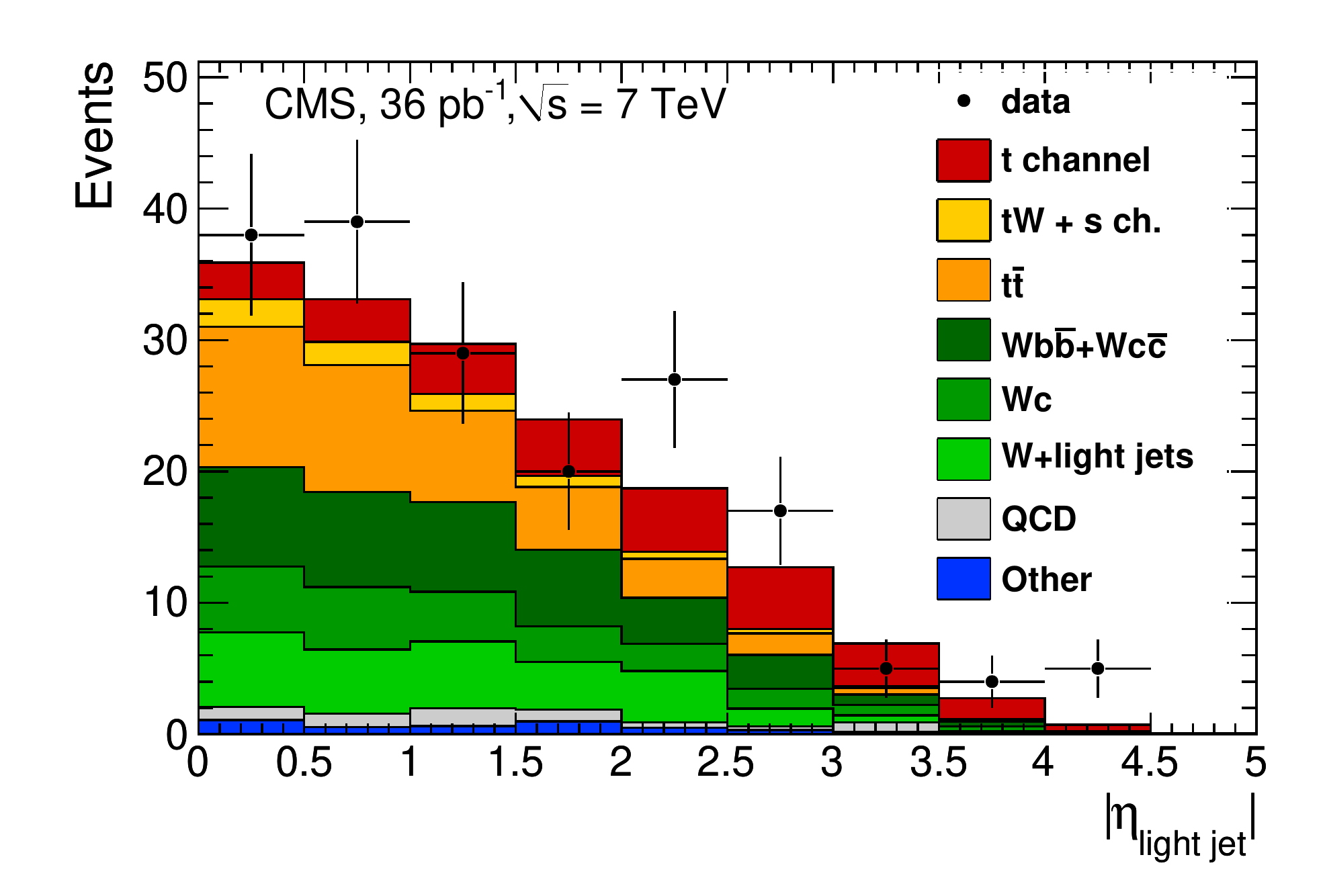}
\caption{Cosine of the angle between charged lepton and untagged jet (\cosThetaPol, top panel) and pseudorapidity of the untagged jet (\etalj, bottom panel) after the 2D selection, for both electron and muon decay channels. QCD and \PW\ + light-partons events are normalized to data, \ttbar, $VQ\bar Q$ ($V=\PW,\cPZ$ and $Q=\cPqb,\cPqc$), and $\PW\cPqc$ are normalized to the result of a dedicated measurement, all other processes are normalized to theoretical expectations.}
\label{fig:2d}
\end{figure}

In the 2D analysis a two-dimensional maximum likelihood fit is performed.
 One of the two fit variables is the cosine of the angle $\theta^*$ between the direction of the outgoing lepton and the spin axis, approximated by the direction of the untagged jet, in the top-quark rest frame~\cite{mahlon1,Motylinski:2009kt}.
 This observable has a distinct slope in signal events, coming from
 the almost 100\% polarization of the top quark due to the $V-A$ structure of the electroweak interaction~\cite{Mahlon:1999gz}.
 This property holds true also in many theories beyond the standard model (SM)~\cite{Batebi:2011kk}.
 The other fit variable is the pseudorapidity distribution of the untagged jet, \etalj, interpreted as the light quark jet recoiling against the single top, whose characteristic $\eta$ distribution allows a discrimination against the typically central jets from the main background processes.
 The distributions in \cosThetaPol and \etalj are shown in Fig.~\ref{fig:2d} for events passing the 2D selection.

 The inputs to the fit are the distributions for signal and backgrounds in the \cosThetaPol-\etalj plane, separately in the muon and electron decay channels.
 The overall background is allowed to float unconstrained in the fit, while its relative components are fixed according to the background estimates. The QCD and \PW\ + light-partons shapes are taken from the anti-isolated and region-$A$ control samples described above, respectively, while all others are taken from the simulation.

\begin{figure}[htpb]
\centering
\includegraphics[width=0.7\columnwidth]{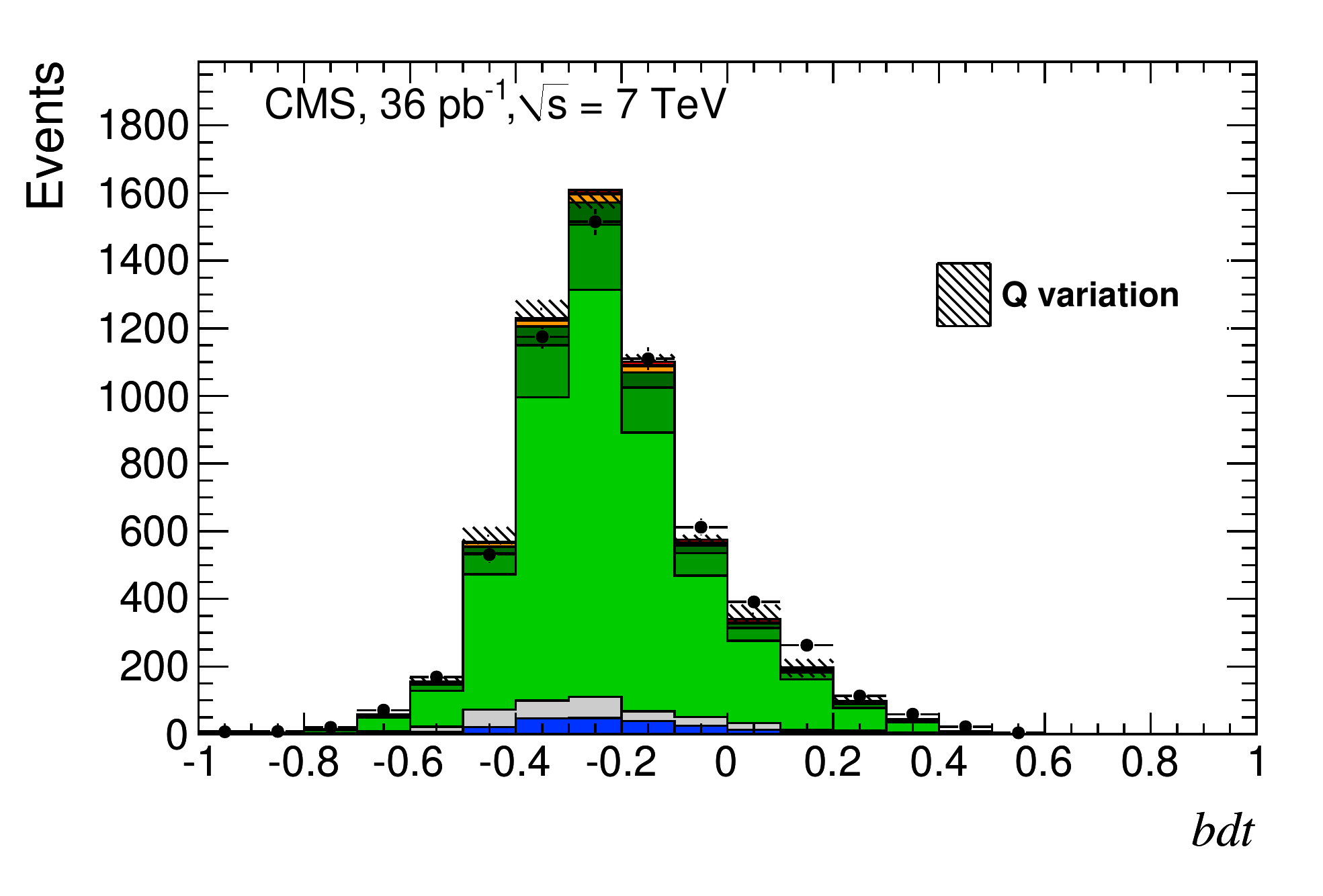}
\includegraphics[width=0.7\columnwidth]{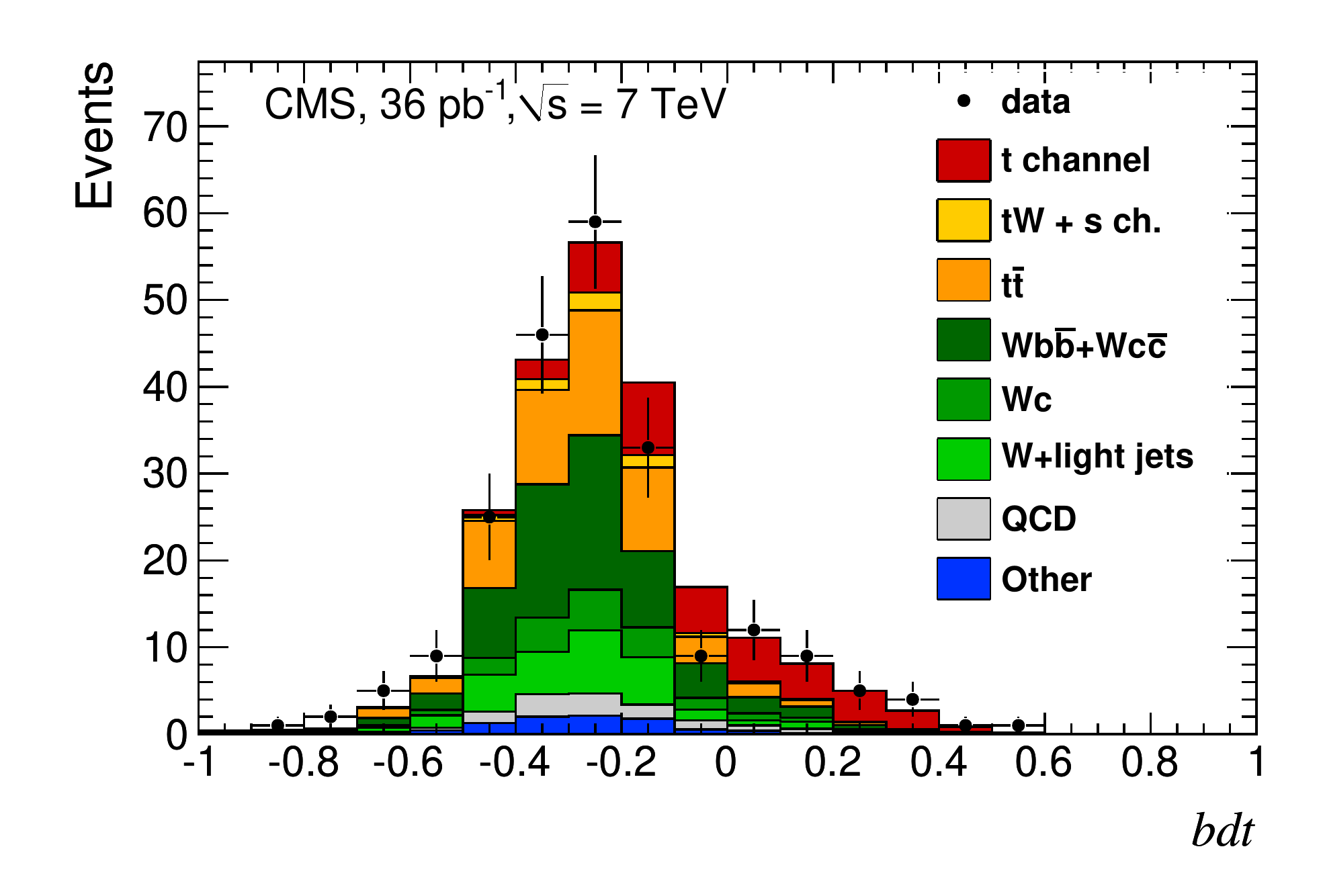}
\caption{Boosted decision tree discriminant ($bdt$) for both electron and muon decay channels in the W-enriched control sample (top panel), with simulation normalized to data, also shown for \PW\ + jets samples with doubled and halved renormalization and factorization scale ($Q$). Same observable after the complete BDT selection (bottom panel), with signal scaled to the measured cross section and all systematic uncertainties and backgrounds scaled to the medians of their posterior distributions.}
\label{fig:bdt_e_mu}
\end{figure}

The BDT method combines a given set of observables into one single classifier variable $bdt$.
 A total of 37 observables have been chosen. Their selection has been inspired by the D0 analysis~\cite{Abazov:2008kt} and optimised for the LHC kinematics.
 The most discriminant ones are the lepton momentum, the mass of the system formed by the reconstructed \PW\ boson and the two jets, the \pt of the system formed by the two jets, the \pt of the jet passing tight \cPqb-tagging requirements, and the reconstructed top-quark mass.
 The validity of the description of all the input variables in the simulation has been checked using a Kolmogorov-Smirnov test in a \PW-enriched control sample with no \cPqb-tagged jet, shown in Fig.~\ref{fig:bdt_e_mu} (top).
 The $bdt$ classifier has been validated both in simulation and in data: negligible differences are found by comparing its distribution for signal events with \MADGRAPH, \textsc{SingleTop}~\cite{boos2000}, and \MCATNLO~3.4~\cite{Frixione:2002ik}, and for $\ttbar$ events with \MADGRAPH, \PYTHIA and \MCATNLO.
 In the \PW-enriched control sample the distribution of $bdt$ from the simulation is statistically compatible with data.

The cross section is extracted from binned $bdt$ distributions using a Bayesian approach.
 The normalizations of the backgrounds
 and the other systematic uncertainties are treated as nuisance parameters.
 The measured distribution of the classifier $bdt$ is shown in Fig.~\ref{fig:bdt_e_mu} (bottom).

The following sources of systematic uncertainties are common to both analyses: background normalization; jet energy scale~\cite{jes}, propagated coherently to the \MET measurement; calibration of the unclustered energy deposits contributing to \MET, varied by $\pm 10\%$; \cPqb-tagging and mistagging efficiencies~\cite{btag}; modeling of the signal and of the main backgrounds; and a 4\% uncertainty on the integrated luminosity~\cite{lumi}.

The uncertainty on the signal model is estimated by comparing \MADGRAPH and \textsc{SingleTop} events with different fragmentation models. The uncertainty on the \ttbar and \PW/\cPZ\ + jets models is determined by comparing simulated samples with varied
 renormalization and factorization scale (within half and double the nominal value, independently for \ttbar and for \PW/\cPZ\ + jets), initial- and final-state radiation parameters, and
 two different fragmentation models.

The impact of pile-up is estimated by comparing the default simulated samples with no pile-up and dedicated samples where minimum bias interactions are superimposed with a probability distribution roughly corresponding to the one observed in the overall 2010 dataset.
 The shapes of the $bdt$ classifier and of both variables used in the 2D analysis are negligibly affected.

In the 2D analysis a conservative systematic uncertainty is assigned to the degree of correlation between \etalj and \cosThetaPol (estimated as 6\% from simulation) by comparing to the result obtained using the product of uncorrelated one-dimensional distributions for the signal.
 The \PW\ + light-partons background shapes in \etalj and \cosThetaPol are extracted from data in the 2D analysis, and studies with simulated events show that the shapes extracted from the control sample are statistically consistent with those in the signal region for the same process. Nevertheless, a small difference is observed in the \etalj shapes in the two selections for the $\PW\cPqc$ process, and we conservatively consider this difference as a systematic uncertainty on all \PW\ + jets processes.

The efficiencies of the muon and electron triggers, identification, and isolation for the 2D selection have been evaluated from data
 using dilepton events at the \cPZ\ peak~\cite{top-10-002}. The uncertainties on these efficiencies have a negligible effect on this analysis.

The impact of each individual source of uncertainty on both analyses has been estimated with an ensemble of pseudoexperiments.
 The dominant systematic uncertainty on the cross section determination comes from the \cPqb-tagging efficiency, known within $\pm 15\%$, because of its large effect on the signal acceptance.
 Nevertheless, this source has a negligible effect on the shapes of the final discriminant variables in both analyses.
 Other important systematic uncertainties come from the signal model, the factorization/renormalization scale for \PW/\cPZ\ + jets, the jet energy scale, and the $\PW\cPqc$ background.

\begin{table}
\caption{Cross section measurements by channel and by analysis. The first uncertainty is statistical, the second systematic. An additional 4\% uncertainty on the luminosity~\cite{lumi} for each measurement is not included.}
\begin{center}
\begin{tabular}{ |c||c|c|}
\hline
Channel         & 2D analysis          & BDT analysis  \\
\hline \hline
$\Pgm$           & $104.1 \pm 42.3 \, _{-28.0}^{+24.8} \,\, \text{pb}$ & $90.4 \pm 35.1 \, _{-19.7}^{+16.5}  \,\, \text{pb}$ \\
$\Pe$       & $154.2 \pm 56.0 \, _{-46.6}^{+40.6} \,\, \text{pb}$ & $59.2 \pm 35.1 \, _{-13.7}^{+13.1}  \,\, \text{pb}$ \\
$\Pgm + \Pe$ & $124.2 \pm 33.8 \, _{-33.9}^{+30.0} \,\, \text{pb}$ & $78.7 \pm 25.4 \, _{-14.6}^{+13.2}  \,\, \text{pb}$ \\
\hline
\end{tabular}
\end{center}
\label{tab:xsec}
\end{table}

Table~\ref{tab:xsec} shows the cross section measured by both analyses in each decay channel, corrected for acceptance and branching ratios. In the muon + electron combination all systematic uncertainties are considered fully correlated, with the exception of the uncertainty on multi-jet QCD obtained from data.
 All measurements are consistent among each other and with the SM expectation.

 Under the assumption that all uncertainties are Gaussian and symmetric, which is fulfilled by the dominant uncertainties,
 the 2D and BDT cross section measurements are combined with the BLUE technique~\cite{Lyons:1988rp}, taking into account a statistical correlation of 51\% estimated with pseudoexperiments, and treating all the systematic uncertainties as fully correlated with the exceptions of those coming from estimates based on data.
 The combined result is
$\sigma^{{\rm exp}} = \xsec \pm \xsecerr \ ({\rm stat.+syst.}) \pm \xseclumi \ ({\rm lumi.}) \,\, \text{pb}$
 where the BDT analysis contributes with the largest weight (89\%).

The expected and observed significances, including systematic uncertainties, are estimated with an ensemble of pseudoexperiments.
 The probability of the predicted background distributions to fluctuate to the observed data corresponds to 3.7 (3.5) Gaussian standard deviations in the 2D (BDT) analysis, combining the electron and muon decay channels, while 2.1$^{+1.0}_{-1.1}$ (2.9$^{+1.0}_{-0.9}$) are expected when assuming SM $t$-channel production cross section.
 The combined significance is well approximated by the BDT significance of 3.5 Gaussian standard deviations.

The single-top cross section measurement can be used as a test of the CKM matrix unitarity~\cite{Alwall:2006bx}
 under the assumption that $|V_{{\cPqt\cPqd}}|$ and $|V_{{\cPqt\cPqs}}|$ are much smaller than $|V_{{\cPqt\cPqb}}|$, and therefore that $|V_{{\cPqt\cPqb}}| = \sqrt{\frac{\sigma^{\text{exp}}}{\sigma^{\text{th}}}}$ where $\sigma^{\text{th}}$ is the SM prediction under the $|V_{{\cPqt\cPqb}}|=1$ assumption.
 Using the prior knowledge that $0\le |V_{{\cPqt\cPqb}}|^2 \le 1$,
 at the 95\% confidence level we infer the lower bound $|V_{{\cPqt\cPqb}}|~>$~0.62 (0.68) from the 2D (BDT) analysis, respectively.

In summary, we confirm the Tevatron observation of single top quark production and present the first measurement of the $t$-channel single top quark production cross section in $\Pp\Pp$ collisions at $\sqrt{s} = 7$~TeV, finding a good agreement with the SM prediction~\cite{Kidonakis:2011wy}.

We wish to congratulate our colleagues in the CERN accelerator departments for the excellent performance of the LHC machine. We thank the technical and administrative staff at CERN and other CMS institutes, and acknowledge support from: FMSR (Austria); FNRS and FWO (Belgium); CNPq, CAPES, FAPERJ, and FAPESP (Brazil); MES (Bulgaria); CERN; CAS, MoST, and NSFC (China); COLCIENCIAS (Colombia); MSES (Croatia); RPF (Cyprus); Academy of Sciences and NICPB (Estonia); Academy of Finland, ME, and HIP (Finland); CEA and CNRS/IN2P3 (France); BMBF, DFG, and HGF (Germany); GSRT (Greece); OTKA and NKTH (Hungary); DAE and DST (India); IPM (Iran); SFI (Ireland); INFN (Italy); NRF and WCU (Korea); LAS (Lithuania); CINVESTAV, CONACYT, SEP, and UASLP-FAI (Mexico); PAEC (Pakistan); SCSR (Poland); FCT (Portugal); JINR (Armenia, Belarus, Georgia, Ukraine, Uzbekistan); MST and MAE (Russia); MSTD (Serbia); MICINN and CPAN (Spain); Swiss Funding Agencies (Switzerland); NSC (Taipei); TUBITAK and TAEK (Turkey); STFC (United Kingdom); DOE and NSF (USA).

\bibliography{auto_generated}   

\cleardoublepage\appendix\section{The CMS Collaboration \label{app:collab}}\begin{sloppypar}\hyphenpenalty=5000\widowpenalty=500\clubpenalty=5000\textbf{Yerevan Physics Institute,  Yerevan,  Armenia}\\*[0pt]
S.~Chatrchyan, V.~Khachatryan, A.M.~Sirunyan, A.~Tumasyan
\vskip\cmsinstskip
\textbf{Institut f\"{u}r Hochenergiephysik der OeAW,  Wien,  Austria}\\*[0pt]
W.~Adam, T.~Bergauer, M.~Dragicevic, J.~Er\"{o}, C.~Fabjan, M.~Friedl, R.~Fr\"{u}hwirth, V.M.~Ghete, J.~Hammer\cmsAuthorMark{1}, S.~H\"{a}nsel, M.~Hoch, N.~H\"{o}rmann, J.~Hrubec, M.~Jeitler, W.~Kiesenhofer, M.~Krammer, D.~Liko, I.~Mikulec, M.~Pernicka, H.~Rohringer, R.~Sch\"{o}fbeck, J.~Strauss, A.~Taurok, F.~Teischinger, P.~Wagner, W.~Waltenberger, G.~Walzel, E.~Widl, C.-E.~Wulz
% ...
\vskip\cmsinstskip
% ...
1:~~Also at CERN, European Organization for Nuclear Research, Geneva, Switzerland\\
% ...

\end{sloppypar}
\end{document}